\documentstyle[12pt,epsf]{article}
\begin{document}
\baselineskip 12pt

\qquad{}

\vskip 1cm

\hrule
\bigskip

\centerline{\large The paper was presented as a poster}

\centerline{\large at JENAM-2000 meeting}

\centerline{\large (29 May - 3 June, Moscow, Russia)}
\bigskip

\hrule
\vskip 2cm

\centerline{\Large {\bf Comparison of photometric parameters}}

\centerline{\Large {\bf of LSB and HSB edge-on galaxies.}}

\vskip 2cm
\centerline{\large \bf Bizyaev D.V.}

\bigskip
\centerline{\large dmbiz@sai.msu.ru}
\bigskip

\centerline{\large Sternberg Astronomical Institute, Moscow, Russia}

\vskip 3cm


\bigskip
\centerline{Abstract}
\bigskip

Photometric parameters of stellar disks and bulges for several edge-on
galaxies from the Catalog of Flat Galaxies (FGC) were determined.
We discuss a difference between photometric parameters of LSB and HSB
galaxies from our sample.

Also we present results of R CCD photometry of edge-on
galaxy RFGC 3647. Deprojecting this galaxy we show that it has
thin LSB disk (central surface brightness $22^m.2$ in R) and
high ratio of radial to vertical scale lengths.
It is shown that
initially gaseous disk of the galaxy was unstable and
its velocity dispersion was low. Stellar disk of this
LSB galaxy was not heated significantly since that time.


\newpage


\medskip
\centerline{\bf Introduction}
\medskip

The main properties of LSB galaxies are
low surface and volume density of stellar disks,
low metallicity and star formation rates and they
are more extended than "normal", HSB galaxies (see de Block etc., 1996,
Bothun etc. 1997, Pickering etc.1994).
The existence of LSB galaxies observed edge-on have remarked in papers
of Karachentsev etc. (1992), in description of Flat Galaxies Catalog
(Karachentsev etc., 1993). In works of Reshetnikov
and Combes (1996, 1997, hereafter RC96, RC97)
small sample of unwarped edge-on galaxies
was investigated and some galaxies of the sample were
suspected to be LSB galaxies.

There was no studies of vertical structure of
a sample LSB galaxies as well as a comparison of photometric
parameters of LSB and HSB spirals obtained from
observations of edge-on objects. We are using the sample of
unperturbed galaxies from papers RC96 and RC97 for these
purposes.

\medskip
\centerline{\bf  The sample of the galaxies.}
\medskip

Images of seven edge-on galaxies obtained in I (Causins)
passband were taken at OHP observatory (France) in July 21-23
1993 by V.Reshetnikov and F.Combes (see RC96 where full description of
CCD-device, processing and flux calibration had been done). Images
were kindly provided us by V.Reshetnikov. Two galaxies of the sample,
UGC 11838 and 11994, have larger surface brightness (hereafter SB)
and belong to 2nd SB class according to FGC,
whereas four of objects, UGC  11241, 11301, 11841 and 11859 have lower SB
and belong to 3rd and 4th SB classes.
Another one galaxy UGC 11132 doesn't included into FGC but
we found the photometric parameters for it because the
inclination of the galaxy is not far from $90^o$.

\medskip
\centerline{\bf  Photometric deprojection. }
\medskip

Photometric parameters of disks and bulges of edge-on galaxies were
obtained by fitting model surfaces to observed 2-D intensity
distributions. The fitting regions for each galaxy were limited
by ellipses oriented according to position angles of the galaxies.
Major axis of the ellipse is chosen to be 0.9 $D_{25}$ and minor
axis equals to 0.1 of the major one (here $D_{25}$ is the optical
diameter). We assumed an exponential low of the distribution
of volume luminosity density for both vertical and radial
directions with scale lengths $Z_e$ and $R_e$ respectively.
Distribution of model surface brightness was found by integration
along the line of sight. Radial extension of the disks is
assumed to be 1.3 $D_{25}$/2 or about four radial scale lengths
(see Zasov etc, 1991). Regions located close to the galactic planes
and central regions were masked to evaluate parameters of the disks.
All objects of our sample look quite symmetric on B, V and I images
so we didn't take into account the difference of the inclinations
from $90^o$.

To fit the model surfaces to the real images of galactic disks
we minimized the sum of squares of residuals changing the parameters
of the model $Z_e$, $R_e$ and central SB. To fit the bulge
we used the result of subtraction best-fit disk model
image from the real one. We used King's function
$I_b (X,Y) = \frac{I_{b0}}{1 + \frac{X^2 + Y^2}{R_{king}^2}}$
to approximate SB distribution in the region of the bulge. Here
$I_{b0}$ and $R_{king}$ are parameters of the model.

\medskip
\centerline{\bf  Results of the deprojection.}
\medskip

Parameters of best-fit models are shown in Table 1. You can
see the name of the object (column 1), distance for
$H_0 ~=~ 75~ km ~s^{-1} Mpc^{-1}$ (2), radial scale length
$R_e$ (3), vertical
scale length $Z_e$ (4), $Z_e / R_e$ relation (5), central
surface brightness of the disks turned face-on in I
passband (6), central SB of the bulges in I (7) and
the class of SB according to FGC (8).

Central SB is found with the help of the calibration
equations taken from RC96 and is corrected to the
extinction in our Galaxy using the information from
LEDA database.

To estimate the errors of photometric parameters
taken by our method we have constructed the sample of
model surfaces. Each model surface was constracted by
summing the exponential disk turned edge-on, the King's
bulge and the noise. It is found that our method
allows us to evaluate the photometric parameters of disks
and bulges with errors not more than 10 \%.
Also we applied our method to the images
of NGC 4244 taken from ING archive (were obtained
at JKT telescope at La Palma) and found
a better agreement between obtained and published
in the paper of van der Kruit and Searle (1981)
values of vertical and radial scales.

The dependence of $R_e$, $Z_e$ $Z_e/R_e$ and $\mu_{0d}$
on SB class is shown at Fig.1. One can see that surface
brightness of disks found for the galaxies of 4th SB class
is about $1.^m5$ lower than for 2nd SB class. It's
well consistent with division of galaxies into LSB and HSB
objects (see for example Tully, Verhejen, 1997, McGaugh, 1996).
Our sample is too small to understand if the scales $R_e$
and $Z_e$ depend on SB class. But relation
$Z_e/R_e$ has a tendency to be lower for the objects
of 4th SB class.

As one can see from Fig.1d, disks of three galaxies of our sample
have low central SB. One of them, UGC 11841, have a scale length
more than 10 Kpc what is tipically for giant LSB galaxies like Malin 1
(see Pickering, Impey, 1996). Note that this galaxy have a minimal value
of $Z_e/R_e$ in our sample. Such value is expected for "flat"
halo-dominated galaxies (Zasov etc., 1991).

Relative magnitudes of the bulges of our galaxies are plotted at
Fig.2a depending on SB class. One can see that the bulges of our LSB galaxies
have larger relative luminosity than the bulges of HSB spirals.
At the same time the central surface brightness of the bulges almost does
not show its dependance on SB class (Fig 2b).

The remnants we got subtracting the model disk and the bulge
from the real image enables us to assume that there are additional components
in the central parts of UGC 11132, 11859, 11994. These galaxies
probably include bars or lenses not considered in our model.

\medskip
\centerline{Table 1.}


\begin{center}
\begin{tabular}{l|rrrrrrrr}
\hline
\hline
Name & D & $R_e$ & $Z_e$ & $Z_e/R_e$ & $\mu_{d0}$ & $\mu_{b0}$ & $R_{king}$ & SB class \\
UGC    & Mpc & kpc & kpc &  & $mag/as^2$ & $mag/as^2$ & kpc & \\
\hline
11132 & 37.9 & 2.8 &  0.67 &  0.24 &  20.9 & 21.7 &  0.3 & -\\
11230 & 94.7 & 8.2 &  1.96 &  0.24 &  21.6 & 23.3 &  0.6 & 4\\
11301 & 60.0 & 8.7 &  1.37 &  0.16 &  20.9 & 23.0 &  0.2 & 3\\
11838 & 46.3 & 4.1 &  0.82 &  0.20 &  21.3 & 26.9 &  0.1 & 2\\
11841 & 79.9 &11.7 &  1.58 &  0.14 &  21.8 & 22.3 &  0.8 & 4\\
11859 & 40.2 & 3.2 &  0.51 &  0.16 &  22.2 & 20.9 &  0.7 & 4\\
11994 & 65.0 & 4.3 &  0.95 &  0.22 &  19.7 & 23.6 &  0.2 & 2\\
\hline
\end{tabular}
\end{center}

\medskip


\medskip
\centerline{\bf  Galaxy RFGC 3647.}
\medskip

We also explored the galaxy RFGC 3647 which images were taken
with better resolution. This object is a good example of edge-on
"thin" galaxy and belongs to 4th SB class according to Revised
Flat Galaxies Catalog (FRGC, Karachentsev etc., 1999). Observations
of the galaxy were held in July 7 , 1999, using 6-m telescope
of Special Astrophysical Observatory. The images were taken
with 300 sec. exposition in Cousin's R passband by I.Karachentsev
and S.Kajsin. Estimated seeing was 1.0" FWHM.
CCD device
Tekram with 1024x1024 pixels provided resolution 0.206 arcsec/pixel.
Images of Landolt's standarts taken at the same night were used for the
calibration purposes. One can see the masked image of the galaxy after
the standard reduction and calibration (Fig. 3).

We used the procedure described before and obtained the photometric
parameters of the disk (radial $R_e$ and
vertical $Z_e$ scales of the disk, both for exponential approximation,
its central surface brightness $\mu_R^d$) and the bulge (central surface
brightness $\mu_R^b$ and core radius $R_{king}$, both for King's bulge).
The main parameters of the galaxy are shown in Table 2. All values of
surface brightness are corrected for the extinction in our Galaxy.

\begin{center}
Table 2. Main parameters of FRGC 3647.
\medskip

\begin{tabular}{|l|r|}
\hline
\hline
R.A.(2000.0) & $20^h ~49^m ~32^s.7$\\
Dec.(2000.0) & $+58^o ~06' ~17"$\\
Distance     & 36.8 Mpc\\
for $H_0=75$ &   \\
&\\
$R_e$        & 5.2 Kpc\\
$Z_e$        & 0.7 Kpc\\
$\mu_R^d$,   & 22.2 $mag/arcsec^2$\\
$Z_e/R_e$    & 0.13\\
&\\
$R_{king}$   & 0.3 Kpc\\
$\mu_R^b$    & 19.8 $mag/arcsec^2$\\
\hline
\end{tabular}
\end{center}
\medskip

As we can see from the Table 2 the central surface brightness of the disk
is about $22^m.2$ in R that is in order of $1.^m5$ lower than its value
for HSB spirals. Total luminosity of the bulge is $2^m$ lower than
the disk's one. So RFGC 3647 should be located close to UGC 11301 on Figures 1
and 2. The radial extension of the disk of RFGC 3647
(limited by the level of S/N = 3) is about 2.7 $R_e$ which is lower than
the tipical value for HSB galaxies.

To investigate how the vertical scale length varies along the radius
of the galaxy we considered few cuts along the minor axis of the galaxy
(see Fig.4). In central parts (R $<$ 2 kpc) the bulge makes $Z_e$ a bit lower
whereas in outer regions mean value of $Z_e \approx 0.66 Kpc$ doesn't change
systematically along the radius. At the large distances from the center
the disk of the galaxy is slightly curved so last points at Fig.4 ($R > 8~
kpc$) shows us increasing of $Z_e$.

The photometric parameters obtained from the observations of edge-on galaxy
enables us to investigate the stability of gaseous disk
before the main event of star sormation (i.e. when the disk was completely
gaseous).

One can compare the critical surface density (according to
the stability criterion
of Polyachenko, see Polyachenko etc. 1997, Zasov, Bizyaev 1996)
and the initial surface density of
the gas (mostly in stars now), see Fig 5a.

Parameters of galactic halo, disk and bulge
obtained by decompisition of the
rotation curve combined with
results of the photometric
deprojection allow us to estimate the velocity dispersion of the
stellar disk, see Fig.5b.
Note that the velocity dispersion of stellar population doesn't decrease
with time and the curve shows an upper limit of velocity dispersion of
initial gaseous disk of the galaxy.
We use artificial rotation curve constracted
by combining the scale length of the disk,
value of maximum of rotative velocity (150 km/s
according to HI-observations by Seeberger etc., 1994) and general form
of rotation corves (Swaters etc., 2000).

As we can see from Fig.5 the
instability criterion may play its role when the galactic disk
forms. The values of velocity dispersion in
gaseous and stellar disks of this LSB galaxy were rather small
(especially in outer regions) compare to lower
limit observed for this value at the present time.



\medskip
\centerline{\bf Conclusions}
\medskip

In the frames of our small sample the galaxies classified as 4th
class of SB in the Flat Galaxies Catalog have
disks with low central surface brightness and relatively
bright bulges.
Deprojecting "thin" edge-on galaxy RFGC 3647 we show that it has
LSB disk (central surface brightness $22^m.2$ in R) and
high ratio of radial to vertical scale lengths.

Surface density of initially gaseous disk was higher than critical
density and gravitational instability slould be an essential
factor affecting the fragmentation of the disk. At the same time
the stellar disk of the galaxy was not heated significantly and its
velocity dispersion is quite low this time.



\vskip 1cm

Acknowlegdements
\medskip

I would like to thank prof. V.Reshetnikov who provided us the images of the
galaxies and prof. I.Karachentsev who allows to process and to publish the
results concerning RFGC 3647. Also I'd like to thank prof. A.Zasov for
usefull discusses and critical notes. We used the NASA/IPAC Extragalactic
Database (NED) and the Lyon/Meudon Extragalactic Database (LEDA) in our
work. The work was supported by grant of Russian Academy of Sciences
(RFBR 98-02-17102).


\bigskip
\centerline{\bf REFERENCES}
\bigskip

\noindent Bothun G., Impey C., McGaugh S. // PASP 1997 v.109.  p.745.

\noindent Gerritsen J., de Blok W. // A.Ap. 1999. v.342. v.655.

\noindent de Blok E., McGaugh S., van der Hulst T. //
BAAS 1996. v.189. 84.02

\noindent Karachentsev I., Georgiev T., Kajsin S. etc.
// Astron. Astrophys. Trans. 1992. v.2. p.265.

\noindent Karachentsev I., Karachentseva V.,
Parnovsky S. // Astron. Nachr. 1993. v.314. p.97. (FGC)

\noindent Karachentsev I.D., Karachentseva V.E., Kudrya Y.N. etc.
// Bull. Special Astrophys. Obs., 1999, v.47, p.5 (RFGC)

\noindent McGaugh S. // MNRAS 1996. v.280. p.337.

\noindent Pickering T., Impey C., van Gorkom J. etc. // BAAS 1994. v.184. 12.08.

\noindent Pickering T., Impey C. // BAAS 1996.  v.186. 39.07.

\noindent Polyachenko V., Polyachenko E., Strelnikov A. // Astronomy
Letters 1997. v.23. p.551.

\noindent Reshetnikov V.,  Combes F.) // A.Ap.Suppl. 1996. v.116.  p.417.
(RC96)

\noindent Reshetnikov V., Combes F. // A.Ap. 1997. v.324.  p.80. (RC97)

\noindent Seeberger R., Huchtmeier W., Weinberger R. // A.Ap. 1994. v.286,
p.17.

\noindent Swaters R., Madore B., Trewhella M. // Ap.J. 2000. v.531.

\noindent Tully R., Verheijen A. // Ap.J. 1997. v.484. p.145.

\noindent van der Kruit P., Searle L. // A.Ap. 1981. v.95. p.105.

\noindent Zasov A., Bizyaev D. // Astronomy Letters 1996. V.22.
P.83.

\noindent Zasov A., Makarov D., Mikhajlova E. // Astr.Letters. 1991, v.17,
p.884

\newpage


\bigskip
\centerline{\bf FIGURES}
\bigskip

\noindent {\bf Fig.1.} The dependence of $R_e$, $Z_e$ $Z_e/R_e$ and $\mu_{0d}$
on the class of surface brightness (according to FGC catalog).

\noindent {\bf Fig.2.} a) Difference between the total magnitudes of bulges and disks
versus the SB class of the galaxies.

b) Central surface brightness of the bulges almost does
not show its dependance on the SB class.

\noindent {\bf Fig.3.} R-image of RFGC 3647 after
the standard reduction, calibration and masking. Surface brightness varies
between 16 and 23 $mag/arcsec^2$ (corrected to the extinction in the Galaxy).
Size of the frame is about 1.5x3.1 arcmin.

\noindent {\bf Fig.4.} Radial behaviour of the vertical scale length
of RFGC 3647.

\noindent {\bf Fig.5.} a) Stellar surface density (solid curve) and critical surface
density (dashed) versus the radial distance.

b) The radial behaviour of the velocity dispersion for totally
gaseous disk estimated using the photometric parameters and
results of decomposition of rotation curve of the galaxy.


\newpage
\begin{center}
\leavevmode
\epsfxsize=16cm
\epsfbox{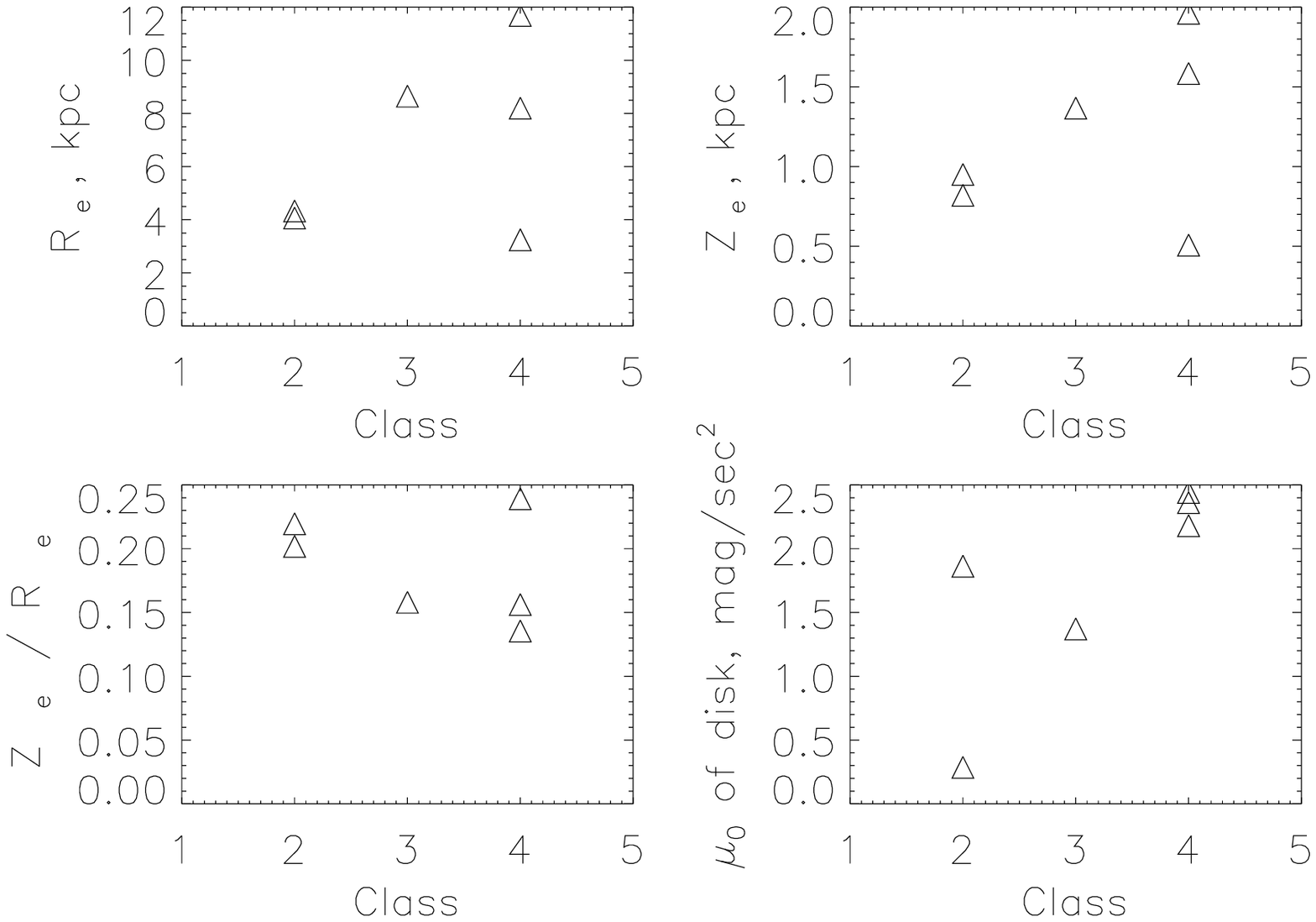}
\end{center}

\newpage
\begin{center}
\leavevmode
\epsfxsize=16cm
\epsfbox{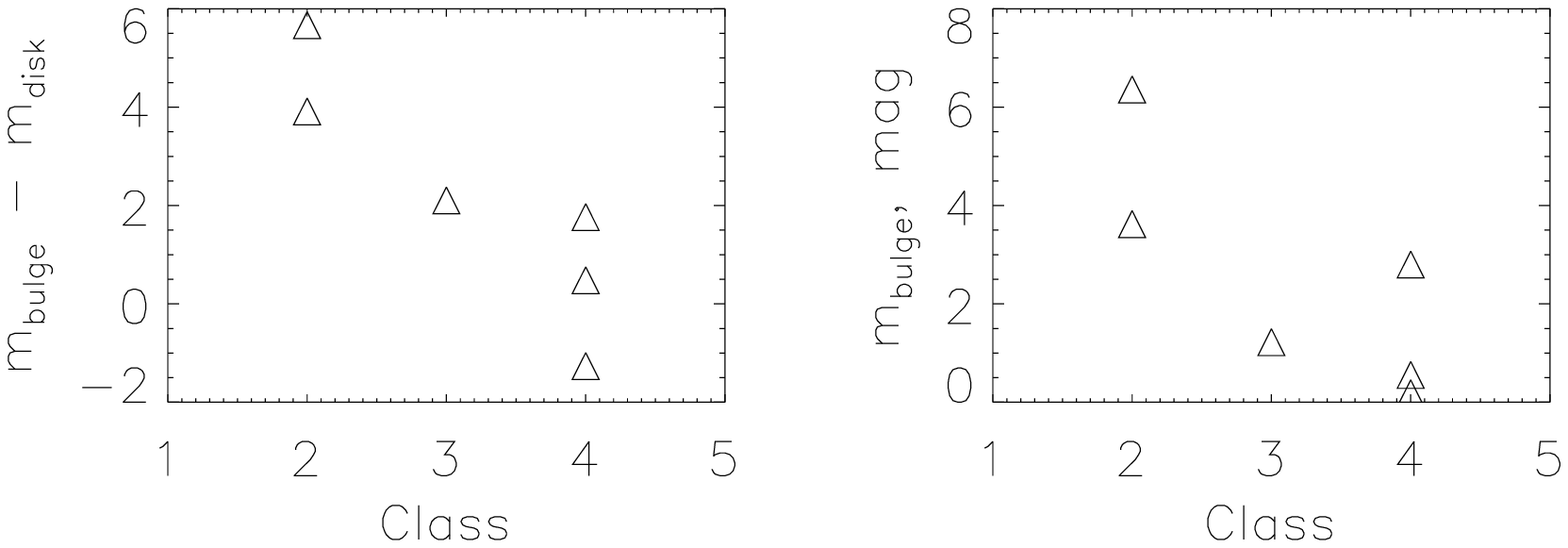}
\end{center}

\newpage
\begin{center}
\leavevmode
\epsfysize=21cm
\epsfbox{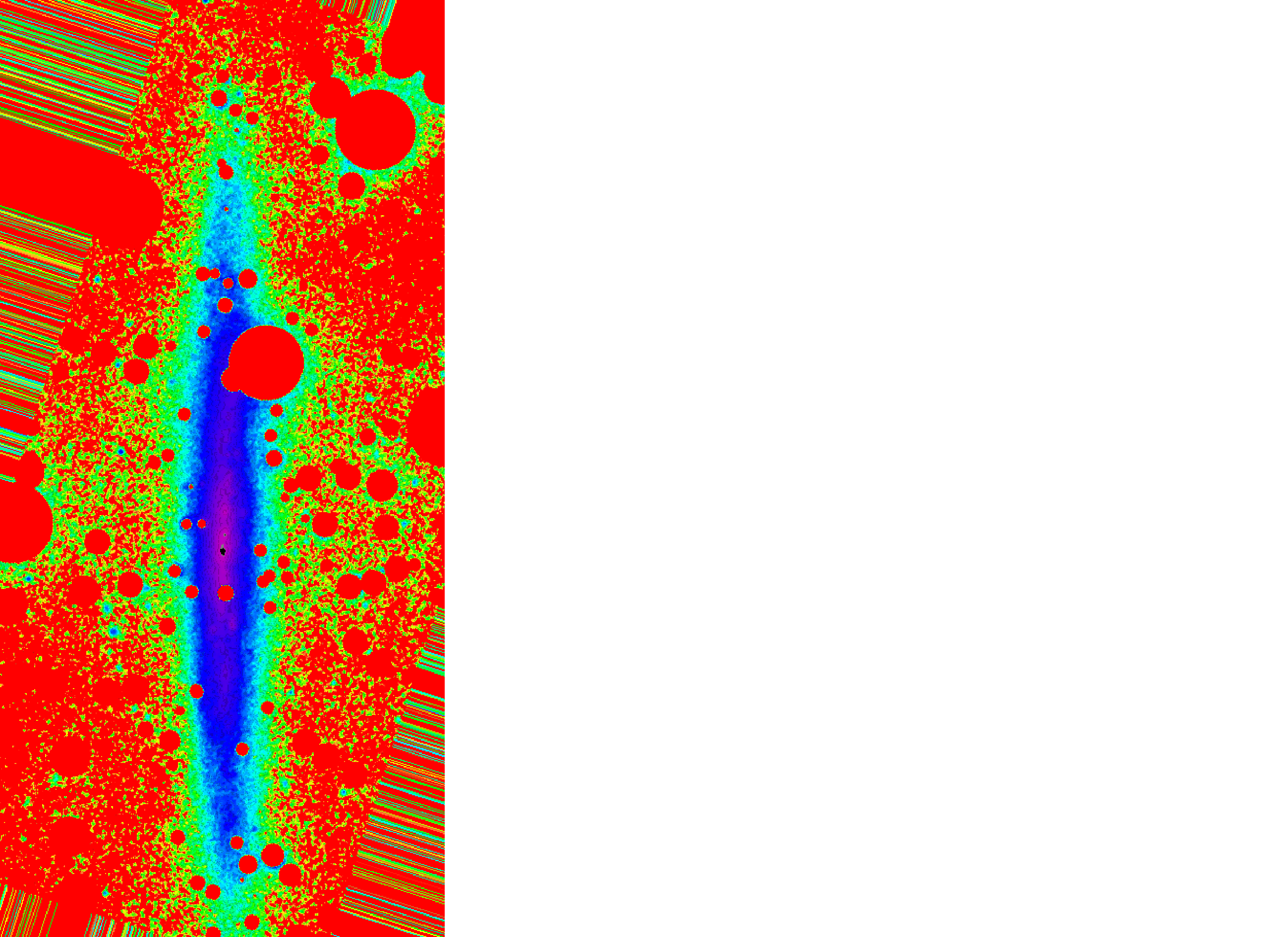}
\end{center}

\newpage
\begin{center}
\leavevmode
\epsfxsize=16cm
\epsfbox{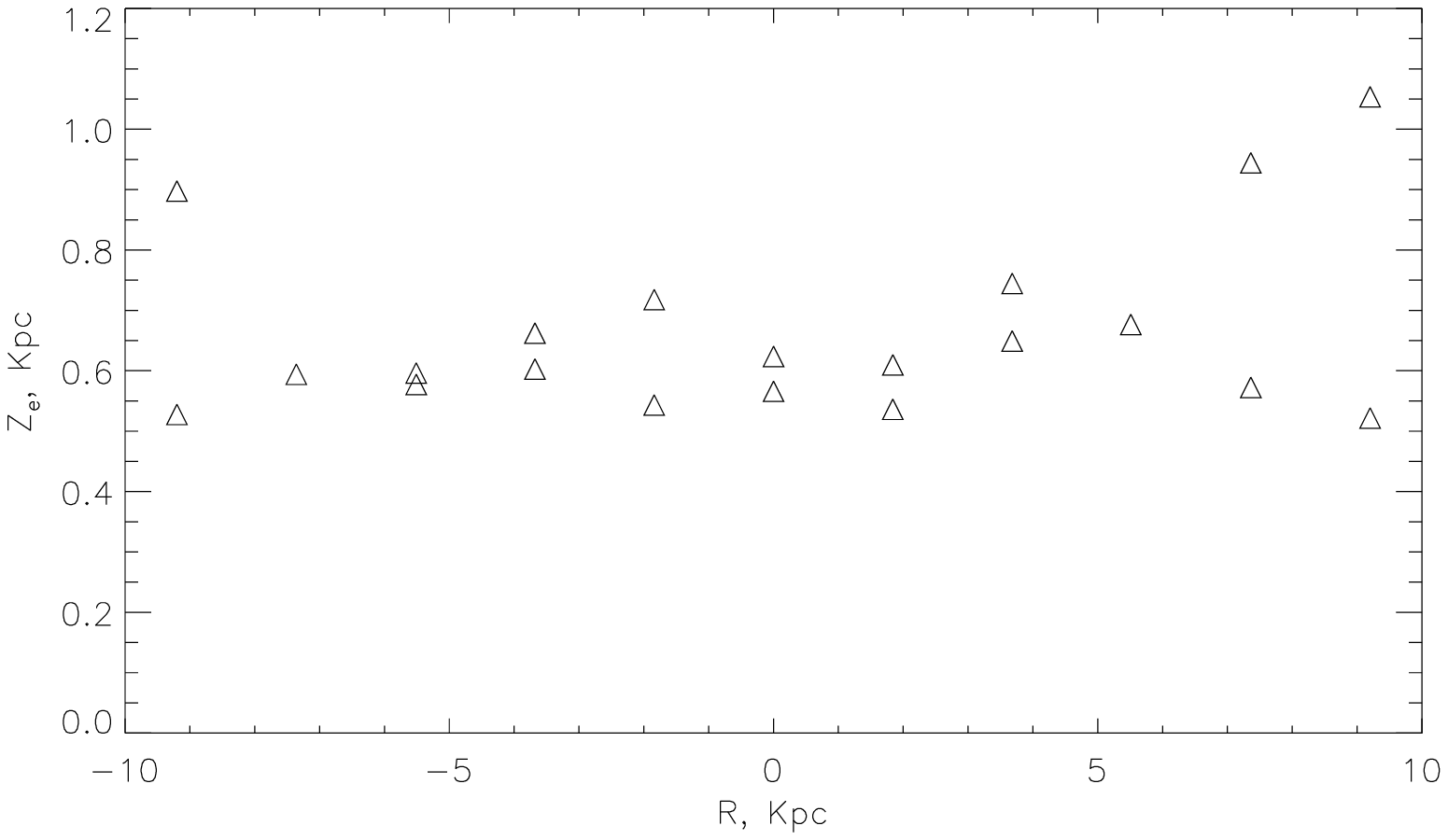}
\end{center}

\newpage
\begin{center}
\leavevmode
\epsfxsize=16cm
\epsfbox{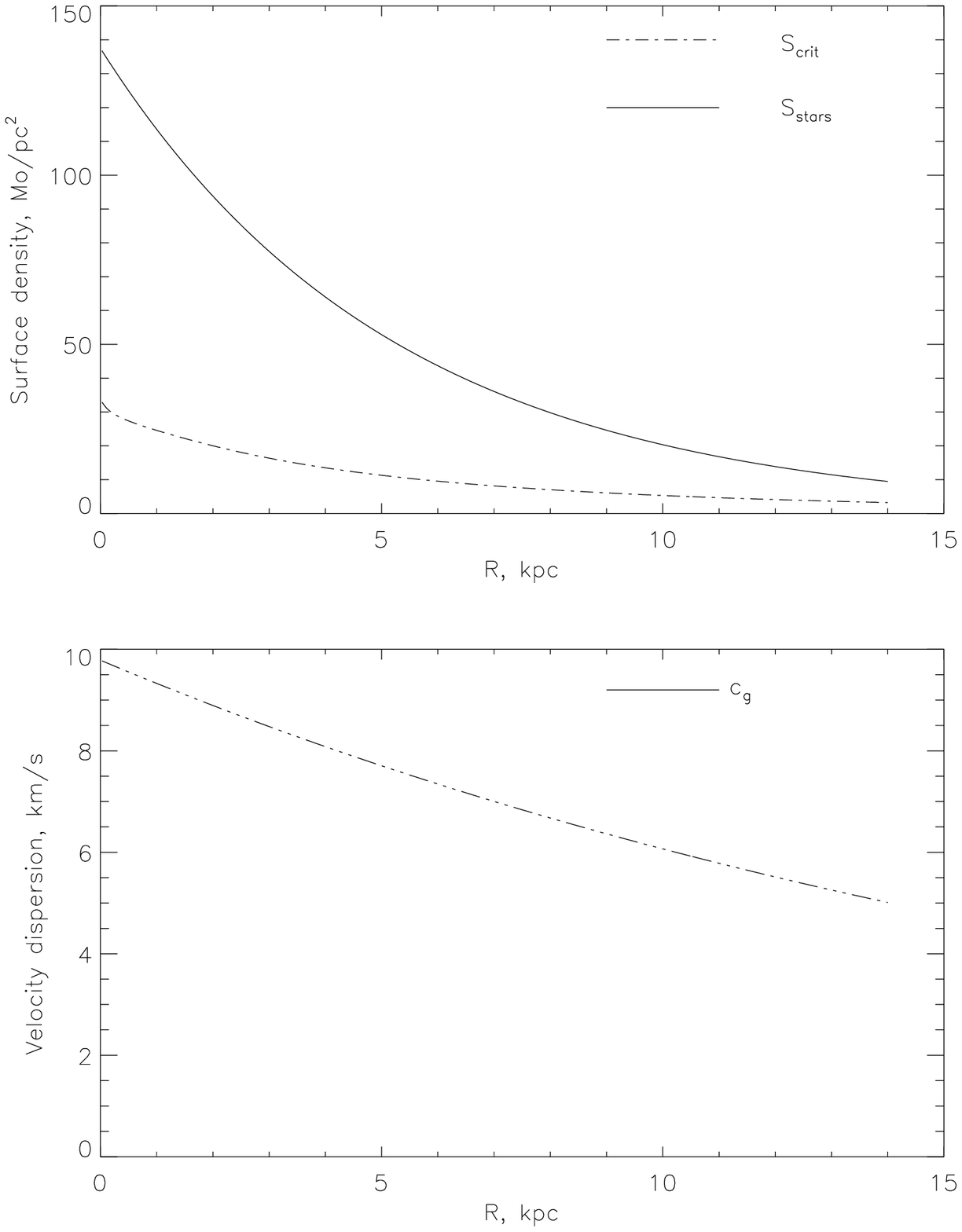}
\end{center}

\end{document}